\documentclass[reprint, amsmath,amssymb,aps, prl]{revtex4-1}

\usepackage{graphicx}
\usepackage{dcolumn}
\usepackage{amssymb,bm}

\usepackage{color}

\newcommand{\Tr}{{\rm Tr}}

\newcommand{\sym}{{\rm sym}}

\begin{document}

\title{Nontrivial UV behavior of rank-4 tensor field models for quantum gravity}

\author{Joseph Ben Geloun}
 \affiliation{ Max Planck Institute for Gravitational Physics, Albert Einstein Institute, Am M\"uhlenberg 1, 14476, Potsdam, Germany\\
 and\\
 International Chair in Mathematical Physics and Applications, ICMPA--UNESCO Chair, 072 B.P. 50  Cotonou, Republic of Benin}
 \email{jbengeloun@aei.mpg.de}

\author{Tim A. Koslowski}
 \affiliation{Instituto de Ciencias Nucleaes, Universidad Nacional Aut\'onoma de M\'exico, Apdo. Postal 70-543, 04510 Ciudad de M\'exico, D.F., M\'excio}
 \email{koslowski@nucleares.unam.mx}

\date{\today}

\begin{abstract}
\noindent 
We  investigate the universality classes of rank-4 colored bipartite $U(1)$ tensor field models near the Gaussian fixed point with the functional renormalization group. In a truncation that contains all power counting relevant and marginal operators, we find a one-dimensional UV attractor that is connected with the Gaussian fixed point. Hence this is first evidence that the model could be asymptotically safe due to a mechanism similar to the one found in the Grosse-Wulkenhaar model, whose UV behavior near the Gaussian fixed point is also described by one-dimensional attractor that contains the Gaussian fixed point. However, the cancellation mechanism that is responsible for the simultaneous vanishing of the beta functions is new to tensor models, i.e. it does not occur in vector or matrix models. 
\end{abstract}

\maketitle

\section{Introduction}

An important breakthrough in quantum gravity research was the discovery of the double- and multicritial scaling limits in matrix models
\cite{Di Francesco:1993nw,Brezin:1990rb,Gross:1989vs}. These CFTs can be interpreted as pure 2-dimensional Euclidean quantum gravity (double scaling limit) coupled to matter (multicritical limits). These scaling limits can be found using techniques of constructive QFT. However, a very useful and unbiased tool for exploring  these scaling limits is the functional renormalization group equation (FRGE) approach to matrix models \cite{Brezin:1992yc,AstridTim,AstridTim2}, where the scaling limits appear as UV fixed points.

It has long been suggested that the success of matrix models in 2 dimensions might be extended to higher dimensions by generalizing matrix models to tensor models \cite{tensor}. Recent years have seen a lot of success in this tensor track program
\cite{Rivasseau:2016wvy,Rivasseau:2011hm,vincentTheorySpace}, starting with the development of the colored (and bipartite) \cite{coloured, tensorNew} specialization of group field theory \cite{boulatov,GFTreviews} supporting 
a large N expansion \cite{largeN} and leading to a new type of universality behavior \cite{criticalTensor,universalityTensor} and which have led to the discovery of perturbative renormalizability of a large class of tensor/group field theories \cite{bengriv,GFTrenorm,mitE, BG,sylvain2,fabdine,Carrozza:2013mna,
BenGeloun:2012pu,josephBeta,Samary:2013xla,Sylvain,Rivasseau:2015ova}. There is thus good motivation to explore the UV structure of colored bipartite tensor field theories with the hope to find attractors of the FRGE flow that one may be able to interpret in terms of quantum gravity.

A different lesson from matrix models was learned in the Grosse-Wulkenhaar model \cite{GW}, which can be viewed as special case of tensor field theories \cite{mitE,BG}. This model was shown to be asymptotically safe near the Gaussian fixed point \cite{Disertori:2006nq}. This asymptotic safety manifests itself in the FRGE approach as a one-dimensional UV attractor that is connected with the Gaussian fixed point \cite{Sfondrini:2010zm}. The connection with the Gaussian fixed point made it technically possible to confirm asymptotic safety purely with FRGE methods in the Grosse-Wulkenhaar model.

Given the a successful application in the description of nonperturbative aspects of quantum Einstein gravity and 
gauge theories \cite{AS,Gies:2006wv},
FRGE methods \cite{Wetterich:1992yh,Delamotte-review,Morris:1993qb} 
have recently been extended to tensor/group field theories with already noteworthy, though preliminary, results (it remains to understand the robustness of the results under enlargement of the truncation and the dependence on the regulator)  \cite{thomasreiko, Krajewski:2015clk, Benedetti:2014qsa, Geloun:2015qfa, Benedetti:2015yaa,Geloun:2016qyb}. Among these results, we cite the confirmation of asymptotic freedom for $\phi^4$-tensor theories \cite{BG,Carrozza:2013mna, BenGeloun:2012pu,Samary:2013xla, Sylvain, Rivasseau:2015ova}, and
the growing evidence of an IR-fixed point which, if confirmed, strongly preludes to the study of phase transitions in such models. 
The different phases could emerge from a symmetry breaking
mechanism leading thereby to the discovery of new vacuum states
(see also how this can be formulated in the context of 
tensor models in \cite{Delepouve:2015nia,Benedetti:2015ara})
and could finally validate the scenario suggesting that
homogeneous and isotropic geometries could emerge
from group field theory \cite{GFTcondensate}. 

The main result of the present letter is that a rank-4, bipartite, colored $U(1)$ tensor field theory also exhibits a one-dimensional UV attractor that is connected with the Gaussian fixed point in a truncation that contains all power counting relevant and marginal operators of the model. The asymptotic safety suggested by this result motivates further an extension of the FRGE investigation of the model to confirm its asymptotic safety. 

Besides the one-dimensional attractor, we found a number of further fixed points whose physical significance was not obvious, and which may very well be truncation artifacts. We thus abstain from the analysis of these fixed points in this letter.

\section{Model and Flow equation}

The specification of a model in the FRGE setup consists of the specification of a theory space and a notion of IR-scale among the elementary degrees of freedom of the model. The elementary degrees of freedom of the model are complex rank-4 tensors $\phi_{\{i_q\}}=\phi_{i_1i_2i_3i_4}$ with colored indices, namely $i_1$ has color $1$ and so on and run from $0$ to $\infty$. Specifically, $\phi_{\{i_q\}}$ can be seen as the Fourier transformed of a complex field $\phi: U(1)^4\to \mathbb{C}$
subjected to a symmetry such that we keep the positive part of the spectrum.  
The geometric interpretation of $\phi_{\{i_q\}}$ is a discrete geometry: for rank 4 it represents a tetrahedron \cite{GFTreviews}.
The IR scale of the model is set by eigenvalues of the Laplacian $\Delta\,\phi_{\{i_q\}}=(i_1^2+i_2^2+i_3^2+i_4^2)\phi_{\{i_q\}}$, so the tensor degrees of freedom for which all indices are small are IR. 

The theory space of the model, i.e. the possible effective average actions $\Gamma_N[\phi]$ that we consider, is the span of all field monomials that satisfy the following rules: (1) the color $j$ index of each tensor $\phi$ is contracted with the same color $j$ index of a complex conjugate tensor $\bar \phi$ and vice versa by a Kronecker delta $\phi_{...i_j...}\delta^{i_j\bar i_j}\bar\phi_{...{\bar i}_j...}$. (2) The only allowed index dependence of the contraction is a non-negative even power of the index, so the contraction has to be of the form $\phi_{...i_j...}(i_j)^{2n}\delta^{i_j\bar i_j}\bar\phi_{...{\bar i}_j...}$ with $n\in \mathbb {N}_o$. 
Moreover, we will assume color-rotation symmetry, i.e. that the effective average action is invariant under relabeling of index colors.
Given that a tensor possesses a discrete 3-geometrical interpretation, interactions may be interpreted as 4-discrete geometries obtained by the gluing 3-geometries along their faces.  

On this theory space we use Wetterich's equation
\begin{equation}
 \partial_t \Gamma_N[\phi] = \overline{\textrm{Tr}}\left((\partial_t R_N)(\Gamma_N^{(2)}[\phi]+R_N)^{-1}\right),
\end{equation}
where $\Gamma_N^{(2)}[\phi]$ denotes the operator obtained by the second variation of the effective average action $\Gamma_N[\phi]$, $\overline{\textrm{Tr}}$ denotes the operator trace and $R_N$ the IR-suppression operator; and where $\partial_t=N\partial_N$ denotes the scale derivative. We will use Litim's optimized cut-off profile \cite{Litim:2001up}  for the IR suppression term, with ${\boldsymbol{\delta}}_{i_q,\bar i_q} = \delta_{i_1\bar i_1}\delta_{i_2\bar i_2}\delta_{i_3\bar i_3}\delta_{i_4\bar i_4}$, 
\begin{equation}
  R_N=Z_N{\boldsymbol{\delta}}_{i_q,\bar i_q}\big(N^2-\sum_{k=1}^4 i_k^2\big) \Theta\big(N^2-\sum_{k=1}^4 i_k^2\big), 
\end{equation}
where $\Theta$ denotes the unit step function. To explicitly evaluate the beta functions, we will use a vertex expansion and identify operators on the RHS of the flow equation analogous to identification in matrix models.

\section{Truncation}

\begin{figure}[h]
     \begin{minipage}{.4\textwidth}
\includegraphics[angle=0, width=5cm, height=1.5cm]{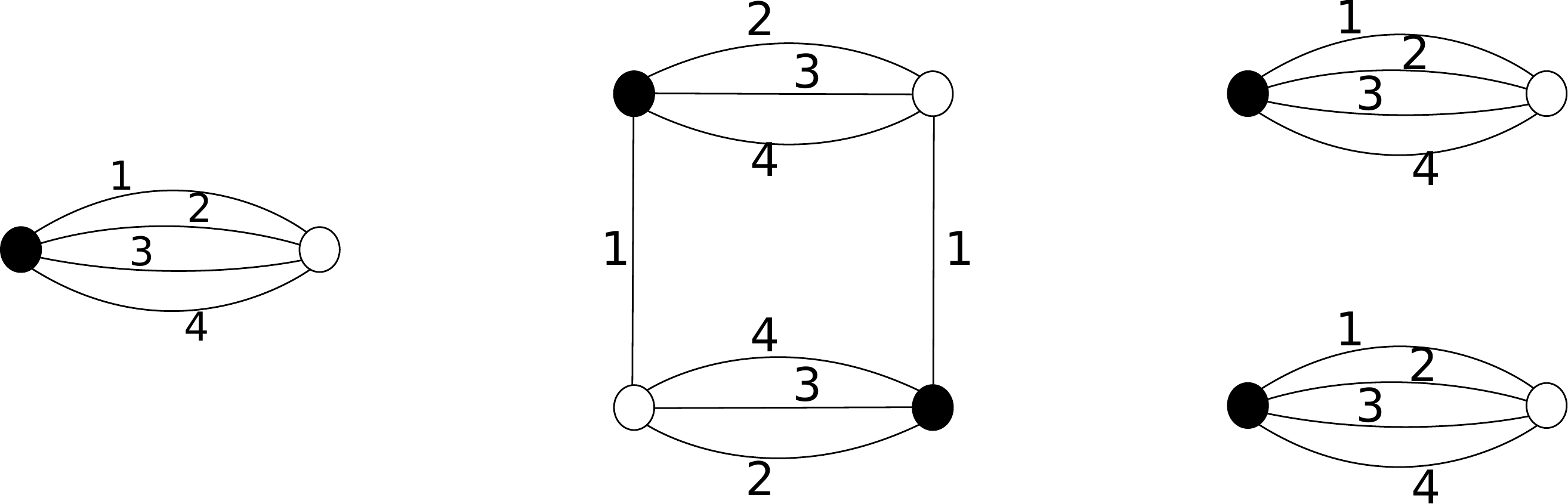} 
\caption{ {\small  The vertices $\Tr(\bar\phi\phi)$, $\Tr_{4;1;1}(\phi^4)$ and $\Tr_{4;2}(\phi^4)$ (respectively, from the left to the right).  }} 
\label{fig:4vertex}
\end{minipage}
\end{figure}

\begin{figure}[h]
\vspace{-0.7cm}
     \begin{minipage}{.5\textwidth}
\includegraphics[angle=0, width=5cm, height=2.1cm]{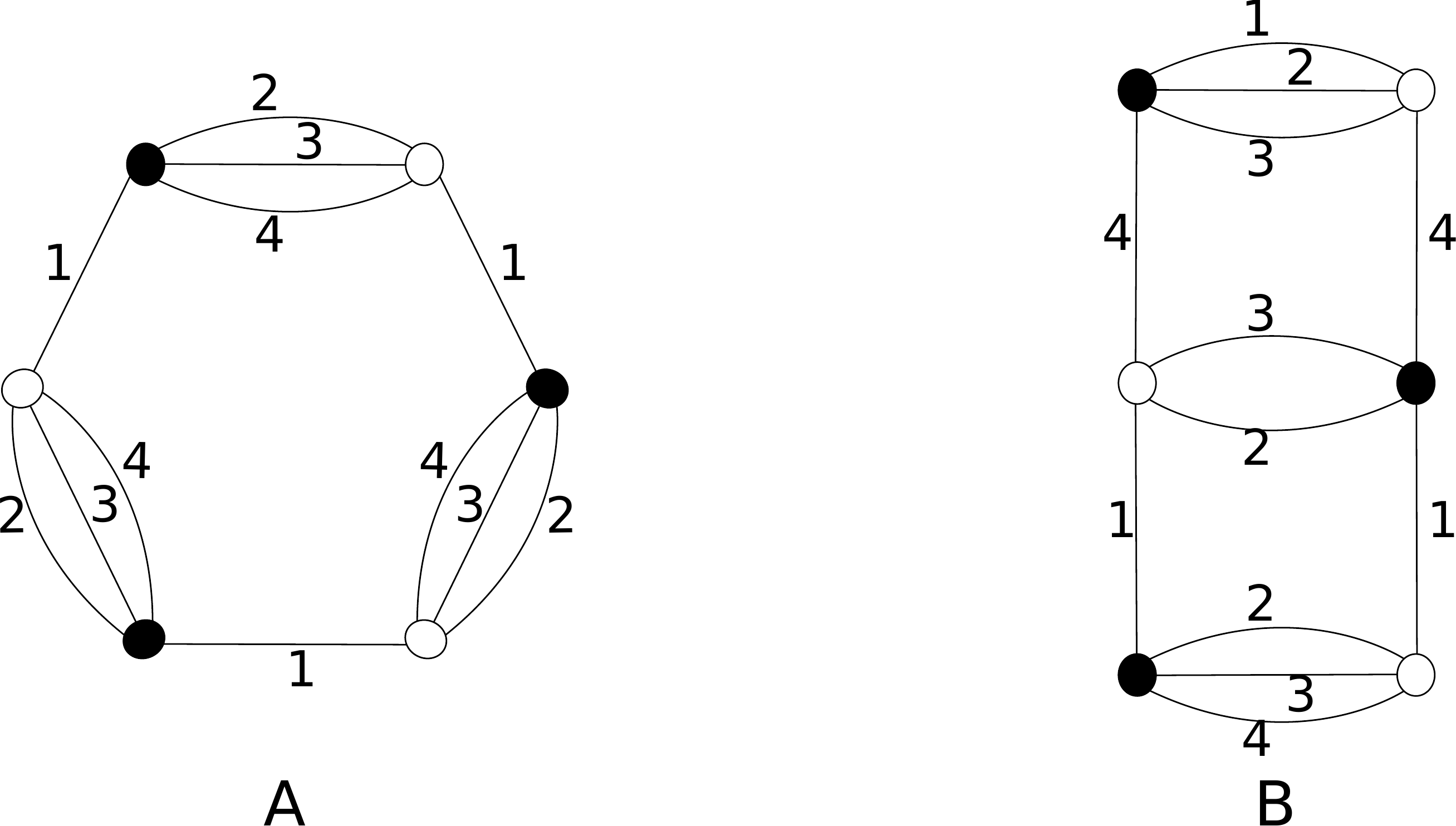} 
\caption{ {\small  The vertices $\Tr_{6;1;1}(\phi^6)$ (A)
and  $\Tr_{6;2;14}(\phi^6)$ (B).  }} 
\label{fig:62vertex}
\end{minipage}
\end{figure}

The beta functions of operators at the Gaussian fixed point are given by the power counting dimension of the operator. This implies that, in the infinitesimal environment of the Gaussian fixed point, the power counting irrelevant operators remain vanishing, since they start with initial value zero and have negative beta functions. Thus, to investigate the infinitesimal environment of the Gaussian fixed point, one needs to consider a truncation that contains all power counting relevant and marginal operators. The power counting for the rank-4 $U(1)$ tensor field theory is known \cite{bengriv} and leads to the following truncation:
\begin{equation}\label{gama}
  \Gamma_N[\phi]=Z_N\,\textrm{Tr}(\bar \phi\,\Delta\,\phi)+m_N\,\textrm{Tr}(\bar\phi\,\phi)+\Gamma_N^{\rm int}[\phi]
\end{equation}
with relevant and marginal interaction terms
\begin{equation}
 \begin{array}{rcl}
 \Gamma_N^{\rm int}[\phi]&=&\frac{\lambda_{4;1}}{2}\, \textrm{Tr}_{4;1}(\phi^4)
+  \frac{\lambda_{4;2}}{2}\, \textrm{Tr}_{4;2}(\phi^4)\\
 &+&
 \frac{\lambda_{6;1}}{3} \, \textrm{Tr}_{6;1}(\phi^6)
+ \lambda_{6;2} \, \textrm{Tr}_{6;2}(\phi^6),
  \end{array}
\end{equation}
where the interaction terms are the symmetrization over colors (denoted by \sym) of the following field monomials (repeated indices are summed)
\begin{eqnarray}
&&
 \textrm{Tr}_{4;1}(\phi^4) = \textrm{Tr}_{4;1;1}(\phi^4)+ \sym \crcr
&&
\textrm{Tr}_{4;1;1}(\phi^4) = 
 \phi_{p_1p_2p_3p_4} \,\bar\phi_{p_{1'}p_2p_3p_4} \,\phi_{p_{1'}p_{2'}p_{3'}p_{4'}} \,\bar\phi_{p_1p_{2'}p_{3'}p_{4'}}\cr\cr
&&
 \textrm{Tr}_{4;2}(\phi^4) =  [\textrm{Tr}_2(\bar\phi\phi)]^2;\;
\textrm{Tr}_2(\bar\phi\phi) = \bar\phi_{p_1p_2p_3p_4} \phi_{p_1p_2p_3p_4}
\cr\cr
&&
 \textrm{Tr}_{6;1}(\phi^6) =   \textrm{Tr}_{6;1;1}(\phi^6)  + \sym \crcr
&&
 \textrm{Tr}_{6;1;1}(\phi^6) = \phi_{p_1p_2p_3p_4} \,\bar\phi_{p_{1'}p_2p_3p_4} \,\phi_{p_{1'}p_{2'}p_{3'}p_{4'}} \crcr
                            && \times \bar\phi_{p_{1''}p_{2'}p_{3'}p_{4'}}\,\phi_{p_{1''}p_{2''}p_{3''}p_{4''}} \,\bar\phi_{p_1p_{2''}p_{3''}p_{4''}} 
\cr\cr
&& \textrm{Tr}_{6;2}(\phi^6) =  \textrm{Tr}_{6;2;14}(\phi^6)
+ \sym \crcr 
&& \textrm{Tr}_{6;2;14}(\phi^6) = \phi_{p_1p_2p_3p_4} \,\bar\phi_{p_{1'}p_2p_3p_4} \,\phi_{p_{1'}p_{2'}p_{3'}p_{4'}} \crcr
                            && \times \bar\phi_{p_1p_{2'}p_{3'}p_{4''}} \,\phi_{p_{1''}p_{2''}p_{3''}p_{4''}} \,\bar\phi_{p_{1''}p_{2''}p_{3''}p_{4'}}  \,. 
\end{eqnarray}
These interactions are graphically represented by bi-partite colored
graphs \cite{tensorNew, Bonzom:2012hw, Gurau:2011tj, Gurau:2012ix}, see figures \ref{fig:4vertex} and \ref{fig:62vertex} and represent particular spherical triangulations.

\section{Beta functions at large $N$}
\label{sect:beta}

The system of beta functions of the model \eqref{gama} is
non-autonomous, that is, it explicitly depends on the cut-off
$N$ even after introducing dimensionless couplings  \cite{Benedetti:2014qsa, Geloun:2015qfa, Benedetti:2015yaa,Geloun:2016qyb}. 
This is the reflection of the existence of 
an external scale \cite{Benedetti:2014gja,Gurau:2009ni}: the radius of the group manifold $U(1)\sim S^1$. 

We are interested in the UV-behavior $N\to\infty$ of the system and so, we restrict to the large $N$-limit of the beta functions which become autonomous. We obtain:
\begin{eqnarray}
&&
 \eta =  \partial_t\ln Z_N = \frac{ 32 a\, g_{4,1;N} }{ (1+ m_N)^2 + 4(1-4a)\,  g_{4,1;N} } \\
&&
 \begin{array}{l}
 \partial_t\, \bar m_N  =
\frac{2}{15(\bar m_N+1)^2
   \left( (1+\bar m_N)^2 + 4(1-4a)\, g_{4,1;N}\right)}   \times\\
 \Big\{ 
-15 (\bar m_N +1)^4\, \bar m_N 
+ 20(\bar m_N +1)^2 \times\\( 3 \bar m_N (8 a-1) + 8 \pi ) g_{4,1;N} \\
-128 (12 a-5)  \pi\, g_{4,1;N}^2
 + 15 \pi ^2 (\bar m_N +1)^2 g_{4,2;N} \\
+20(3-8a) \pi ^2\,   g_{4,1;N} g_{4,2;N}
\Big\}
 \end{array}
\\
&&
\begin{array}{l}
   \partial_t \, g_{4,1;N}  = \frac{1}{15(\bar m_N+1)^3
 \left( (1+\bar m_N)^2 + 4(1-4a)\, g_{4,1;N}\right)} \times\\
\Big\{ 
 (-15 (1 + \bar m_N)^4 
+(480 -1152 a)\,\pi (g_{6,1;N}+g_{6,2;N} ))\\ \times(1 +\bar m_N) g_{4,1;N} \\
+ (1 +\bar m_N)^2 ( (1200a-60) (1 +\bar m_N) - 160\pi) 
g_{4,1;N}^2 \\
 +(-640+1536 a)\, \pi g_{4,1;N}^3
+ 120\pi (1+\bar m_N )^3 \\(g_{6,1;N}+g_{6,2;N} )
\Big\}
 \end{array}
\\
&&
 \begin{array}{l}
\partial_t \,g_{4,2;N}  = \frac{4}{15(\bar m_N+1)^3
 \left( (1+\bar m_N)^2 + 4(1-4a)\, g_{4,1;N}\right)} \times  \\
\Big\{ 
-360 \pi (1 + \bar m_N)^2 g_{4,1;N}^2 +  1440(2a-1) \pi\, g_{4,1;N}^3\\
+
(1 + \bar m_N)^2  (-1280\pi  + 
   672 (1 + \bar m_N) ) g_{4,1;N}g_{4,2;N} \\
+ (320 a -120 ) \,\pi^2\, g_{4,1;N}g_{4,2;N}^2 
+  (-1280 + 3072 a) \\ \times \pi\,  g_{4,1;N}^2g_{4,2;N}  \\
-30\pi^2\,(1 + \bar m_N)^2 g_{4,2;N}^2
+ \pi (1 + \bar m_N) (45 (1 + \bar m_N)^2 \\ + (180  - 360 a) \,g_{4,1;N} ) g_{6,2;N} ) \Big\}
 \end{array}
\\
&&
 \begin{array}{l}
\partial_t\, g_{6,1;N}  =\frac{32}{15(\bar m_N+1)^4
 \left( (1+\bar m_N)^2 + 4(1-4a)\, g_{4,1;N}\right)} \times \\
g_{4,1;N} \Big\{
\pi ( (96a-40)\,  g_{4,1;N} - 10  (1+\bar m_N)^2)
g_{4,1;N}^2  \\
+ 
\left(   (1+\bar m_N)^3 ( 45 a  (1+\bar m_N) - 15  \pi ) 
+ \pi(1+\bar m_N)\right. \\ \left. ( 144 a- 60  ) g_{4,1;N}  \right) g_{6,1;N} \\
 (1+\bar m_N)^3 
(1008 (1+\bar m_N) - 480 \pi ) g_{6,1;N}\\
+ 1305.6 \pi (1+\bar m_N) g_{4,1;N} 
\Big\}
 \end{array}
\\
&&
 \begin{array}{l}
    \partial_t \, g_{6,2;N}  =\\
\hspace{-0.2cm}
\frac{32 \left( 45 a (1+\bar m_N)^3 - 10 \pi (1+\bar m_N)^2
+ \pi (96a-40) g_{4,1;N} \right)g_{4,1;N} g_{6,2;N} }{15(\bar m_N+1)^3
 \left( (1+\bar m_N)^2 + 4(1-4a)\, g_{4,1;N}\right)}
 \end{array}
\end{eqnarray}
The spectral sums on the RHS of the flow equation contain index dependent sums which were not known to us analytically, so we  evaluated them  numerically. This leads to the appearance of a numerical constant $a\in [0.7,0.8]$ in the system of beta functions. In the following, we use
$a = 0.7$
and stress that the results are qualitatively the same for several other
values in the full range of $a$.

\section{UV Attractor}

We now look at the fixed point condition, i.e. the condition for the simultaneous vanishing of all beta functions in the previous section. Besides a number of isolated fixed points (whose physical significance we can not judge in the present truncation), we find a one-dimensional attractor going through the Gaussian fixed point that is characterized by $g_{4,1}=0$ and $g_{6,1}=-g_{6,2}$. This second condition is genuinely new to tensor models (i.e. it does not appear in matrix models) since it corresponds to a cancellation between the effects of the two power counting marginal six point interactions. The UV attractor can be parametrized by a parameter $\lambda$:
\begin{equation}
 \begin{array}{l}
  \left(m^*,g^*_{4,1},g^*_{4,2},g^*_{6,1},g^*_{6,2}\right)(\lambda)\,=\quad\\
 \quad\quad\left(\lambda,0,\frac{\lambda (\lambda+1)^2}{\pi ^2},-\frac{2 \lambda^2 (\lambda+1)^3}{3 \,\pi ^3},\frac{2 \lambda^2 (\lambda+1)^3}{3\, \pi ^3}\right)
  \end{array}
\end{equation}
We trust our truncation near the Gaussian fixed point and the vertex expansion as long as we are far away from $m^*=-1$. Near the Gaussian fixed point, and for $\lambda> 0$, we have $g^*_{6,1}<0$, so it is unclear whether the action is bounded from below or not, since this depends on the relative growing of the two marginal six point monomials. 

There is no flow tangent to the attractor, i.e. the direction
\begin{equation}
 \begin{array}{l}
   \left(dm^*,dg^*_{4,1},dg^*_{4,2},dg^*_{6,1},dg^*_{6,2}\right)=\\
\left(1,0,\frac{(\lambda+1) (3 \lambda+1)}{10 \pi ^2},-\frac{2 \lambda (\lambda+1)^2 (5 \lambda+2)}{3 \pi ^3},\frac{2 \lambda (\lambda+1)^2 (5 \lambda+2)}{3 \pi ^3}\right)
 \end{array}
\end{equation} 
at a fixed point parametrized by $\lambda$ is a completely marginal deformation of the fixed point. The attractor contains the Gaussian fixed point, which is attained at $\lambda=0$ in our parametrization. This is analogous to the Grosse-Wulkenhaar model, which also contains a line of fixed points that emanates from the Gaussian fixed point.

The Hessian at the attractor depends on the parameter $\lambda$
and is given by the matrix $\partial_{g_j} \beta_i$ of the form
\begin{equation}
\left[
\begin{array}{ccccc}
 \frac{-2(1+3\lambda)}{\lambda +1} & \frac{4 (b \lambda +16 \pi )}{3 (\lambda +1)^2} & \frac{2 \pi ^2}{(\lambda +1)^2} & 0 & 0 \\
 0 & -1 & 0 & \frac{8 \pi }{(\lambda +1)^2} & \frac{8 \pi }{(\lambda +1)^2} \\
 \frac{8 \lambda ^2}{\pi ^2} & \frac{2 \lambda  (b (4 \lambda +3)-128 \pi )}{3 \pi ^2 (\lambda +1)} & -\frac{16 \lambda }{\lambda +1} & 0 & \frac{12 \pi }{(\lambda +1)^2} \\
 0 & \frac{2 \lambda ^2 (32 \pi -3 b (\lambda +1))}{3 \pi ^3} & 0 & 0 & 0 \\
 0 & \frac{2 \lambda ^2 (9 b (\lambda +1)-64 \pi )}{9 \pi ^3} & 0 & 0 & 0 \\
\end{array}
\right]
\end{equation}
where  $b=22.4$. The analytic expression for the five eigenvalues $\theta_1$, $\theta_2^\pm$ and $\theta_3^\pm$ are:
\begin{eqnarray} 
 \theta_1&=&0\\
 \theta_2^\pm&=&\frac{1}{\lambda+1}\left((-11\lambda+1)\pm\sqrt{1+\lambda(41\lambda+10)}\right)\\
 \theta_3^\pm&=&-\frac 1 2 \pm \frac{\sqrt{2048\lambda^2+9\pi(\lambda+1)^2}}{6(\lambda+1)\sqrt{\pi}}
\end{eqnarray}
The exactly vanishing critical exponent $\theta_1$ is associated with the tangent direction to the attractor. The eigenvalues $\theta_2^-$ and $\theta_3^-$ are negative for all $\lambda>0$ and the associated eigen-direction span essentially the mass and the power-counting relevant coupling $g_{4,1}$ in the vicinity of the Gaussian fixed point, where they attain the canonical values $-2$ and $-1$. The remaining two critical exponents $\theta_2^+$ and $\theta_3^+$ vanish at the Gaussian fixed point. It follows that the critical exponents $\theta_1$, $\theta_2^+$ and $\theta_3^+$ correspond to the power counting marginal couplings at the Gaussian fixed point.
\begin{figure}
\includegraphics[angle=0, width=5.5cm, height=6cm]{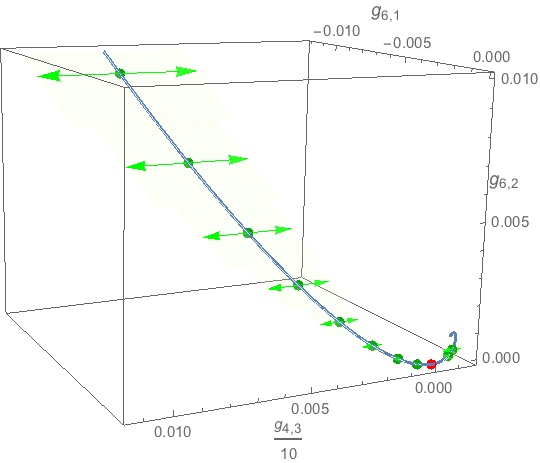}
  \caption{Projection of the attractor and its critical surface onto the space of marginal couplings. The Gaussian fixed point is indicated in red, the attractor as the blue curve and the critical surface (as well as the restriction of the flow to the critical surface) in light green.}
 \label{fig:CriticalSurface}
 \end{figure}
For $\lambda>0$, i.e. away from the Gaussian fixed point, we find that $\theta_2^+$ becomes negative, while $\theta_3^+$ becomes positive. This means that the eigen-direction associated with $\theta_2^+$ is a relevant deformation of the attractor that turns marginal in the Gaussian limit $\lambda\to 0$, while the eigen-direction associated with $\theta_3^+$ is an irrelevant deformation of the attractor, which also turns marginal in the Gaussian limit.  
 
We can thus understand the critical surface associated with the attractor. Since we are interested in the Gaussian fixed point, we concern ourselves with the power counting relevant directions $m$ and $g_{4,1}$ which are turned on by the flow. We  can thus project the flow onto the three-dimensional space of power-counting marginal couplings, which we depict in figure \ref{fig:CriticalSurface}. The figure is explained as follows: The blue curve indicates the location of the attractor, which contains the Gaussian fixed point (red dot). The light green surface indicates the critical surface associated with this attractor and the arrows on the critical surface indicate the flow towards the IR (i.e. the relevant deformations of the attractor). The length of the green arrows indicates the strength of the flow. This shows graphically that the relevant deformation turns marginal at the Gaussian fixed point.

\section{Conclusion}

In this letter we investigated a bipartite, colored $U(1)$ tensor field theory of rank-4 with color permutation symmetry with the functional renormalization group equation. We are in particular interested in the UV-behavior of the system very close to the Gaussian fixed point. We thus consider a truncation that contains all power counting relevant and marginal operators of the theory space. In this truncation we find a one-dimensional attractor that is connected with the Gaussian fixed point. This attractor is due to a cancellation of the contributions of the two power counting marginal six point interactions of the tensor model and is thus a genuinely new effect that can as such not appear in matrix models. 

This one-dimensional attractor provides evidence that the rank-4 tensor model is asymptotically safe in a manner analogous to the Grosse-Wulkenhaar model, where asymptotic safety is found as a one-dimensional UV attractor that contains the Gaussian fixed point. Since rank-4 tensor models where developed as models for 4 dimensional Euclidean quantum gravity, this evidence for asymptotic safety and the genuinely new cancellation mechanism that causes it, could have important consequences for quantum gravity. 

The present work therefore corrects the conclusion of the perturbative flow analysis of 
the very same rank-$4$ $\phi^6$ model in a previous paper \cite{josephBeta} (partly also reported in \cite{BG}) claiming that the model is asymptotically free in the UV. The perturbative calculations in that reference were performed at 4-loop order, due a cancellation of the beta function in one sector (i.e. $g_{6;1}$) at first loops (already providing strong evidence of asymptotic safety). However, 
the unclear use of coupling scaling dimensions and 
the approximations made in that paper to push the calculations to higher order 
loops, cast obviously doubt on their validity and on the system of beta functions.  At the same time, this work corroborates (with the well understood limits of FRGE truncations) asymptotic safety for $\phi^6$-models  predicted  in the group field theory context \cite{Sylvain}. In the Gaussian limit of renormalizable tensor theories, one expects \cite{Rivasseau:2015ova}, that the wave function renormalization tends to dominate the coupling renormalization, making 
the model either asymptotically free or asymptotically safe. 

\paragraph{Acknowledgements:} TK thanks the
Max-Planck Institute for Gravitational Physics  (Albert Einstein Institute),  Potsdam, for its hospitality while doing significant parts of the work for the present letter.

\end{document}